\documentclass[9pt, twocolumn]{article}
\usepackage{times,bm}
\usepackage[affil-it,]{authblk}
\usepackage[pagebackref=false,colorlinks=true,linkcolor=olive,citecolor=red,urlcolor=acsblue]{hyperref}
\usepackage{geometry}
\usepackage[switch*, displaymath,pagewise, mathlines]{lineno}
\usepackage{graphicx}
\usepackage{cite}
\usepackage{amsmath}
\usepackage{amsfonts}
\usepackage{amssymb}
\usepackage{slashed}
\usepackage{subfloat}
\usepackage{slashed}
\usepackage{color}
\usepackage{float}
\usepackage{multirow,multicol}
\usepackage{tabulary}
\usepackage[flushleft]{threeparttable}
\usepackage{pgfplots,subfigure}
\usepackage{xcolor}
\numberwithin{equation}{section}
\usepackage{tikz}
\usepackage{ulem}
\usepackage{hyperref}
\usepackage{nameref}
\usepackage{comment}
\usepackage{xcolor}
\definecolor{acsblue}{RGB}{17,76,139}
\definecolor{shadecolor}{RGB}{255,241,204}

\usepgflibrary{arrows}
\usetikzlibrary{shapes.callouts}
\tikzset{
	level/.style   = { thick, },
	connect/.style = { dotted, red   },
	notice/.style  = { draw, rectangle callout, callout relative pointer={#1} },
	label/.style   = { text width=2cm }
}
\usepackage{letltxmacro}
\makeatletter
\let\oldr@@t\r@@t
\def\r@@t#1#2{%
	\setbox0=\hbox{$\oldr@@t#1{#2\,}$}\dimen0=\ht0
	\advance\dimen0-0.2\ht0
	\setbox2=\hbox{\vrule height\ht0 depth -\dimen0}%
	{\box0\lower0.4pt\box2}}
\LetLtxMacro{\oldsqrt}{\sqrt}
\renewcommand*{\sqrt}[2][\ ]{\oldsqrt[#1]{#2}}
\makeatother
\usepackage{environ}

\begin{document}

\def\nofundrefquery{}
\def\stmdocstextcolor#1{}
\def\stmdocscolor#1{}
\newcommand{{\ri}}{{\rm{i}}}
\newcommand{{\Psibar}}{{\bar{\Psi}}}
\renewcommand{\rmdefault}{ptm}

\fontsize{8}{9}\selectfont

\title{\mdseries{Twist-Induced Effects on Weyl Pairs in Magnetized Graphene Nanoribbons}}

\author{ \textit {\mdseries{Semra Gurtas Dogan$^{\textbf{*}}$}}$^{\ 1}$\footnote{\textit{ E-mail: semragurtasdogan@hakkari.edu.tr (Corr. Auth. $^{\textbf{*}}$ )} }~,~ \textit {\mdseries{Kobra Hasanirokh}}$^{\ 2}$\footnote{\textit{ E-mail: zhasanirokh@yahoo.com } }~,~ \textit {\mdseries{Omar Mustafa}}$^{\ 3}$\footnote{\textit{ E-mail: omar.mustafa@emu.edu.tr} }~,~ \textit {\mdseries{Abdullah Guvendi}}$^{\ 4}$\footnote{\textit{ E-mail: abdullah.guvendi@erzurum.edu.tr } }  \\
	\small \textit {$^{\ 1}$ \footnotesize Department of Medical Imaging Techniques, Hakkari University, 30000, Hakkari, Türkiye}\\
	\small \textit {$^{\ 2}$ \footnotesize Research Institute for Applied Physics and Astronomy, University of Tabriz, Tabriz 51665-163, Iran}\\
	\small \textit {$^{\ 3}$ \footnotesize Department of Physics, Eastern Mediterranean University, 99628, G. Magusa, north Cyprus, Mersin 10 - Türkiye}\\
 \small \textit {$^{\ 4}$ \footnotesize Department of Basic Sciences, Erzurum Technical University, 25050, Erzurum, Türkiye}}
\date{}
\maketitle

\begin{abstract}
This paper presents an analytical investigation into the dynamics of Weyl pairs within magnetized helicoidal graphene nanoribbons. By embedding a curved surface into flat Minkowski space-time, we derive a fully covariant two-body Dirac equation specific to this system. We begin by formulating a non-perturbative wave equation that governs the relative motion of the Weyl pairs and obtain exact solutions. Our results demonstrate the influence of the uniform magnetic field and the number of twists on the dynamics of Weyl pairs in graphene nanoribbons, providing precise energy values that lay a robust foundation for future research. Furthermore, we examine the material's response to perturbation fields by calculating the polarization function and investigating how twisting and magnetic fields affect this response. Our findings indicate that, in principle, the material's properties, which are crucial for practical applications, can be effectively controlled by precisely tuning the magnetic field and the number of twists in graphene nanoribbons.
\end{abstract}

\begin{small}
\begin{center}
\end{center}
\end{small}

%\bigskip

%\maketitle

\section{\mdseries{Introduction}}\label{sec1}

Graphene \cite{geim2007rise}, a single layer of carbon atoms arranged in a hexagonal lattice, has revolutionized materials science and condensed matter physics due to its extraordinary electronic, mechanical, and thermal properties. Its high electrical conductivity, mechanical strength, and thermal conductivity have made graphene a material of immense interest across various fields \cite{castro2009electronic}. When graphene is shaped into different geometries, such as nanoribbons (NRs), and subjected to external influences like magnetic fields, its electronic behavior can become highly complex \cite{deng2016wrinkled}. Twisted graphene nanoribbons (TGNRs), formed by applying a twist to the graphene lattice, exhibit unique electronic and structural properties. This twist alters the material's electronic band structure, leading to the emergence of new electronic states. The interaction between these twisted structures and external magnetic fields can further modify their electronic properties, providing fertile ground for discovering novel physical phenomena \cite{kwan2020twisted,background}. In the presence of a uniform magnetic field, charge carriers in these materials are significantly influenced by the Lorentz force, resulting in phenomena such as the quantum Hall effect and the formation of Landau levels \cite{zhang2005experimental}. The twist in the graphene lattice introduces a periodic potential, which can lead to the formation of Weyl pairs that behave as massless entities with chirality. These particles (massless fermion-antifermion ($f\overline{f}$)  pairs) in TGNRs are of particular interest due to their unique dispersion relations and the potential for observing chiral anomaly effects and topological phenomena \cite{armitage2018weyl}. Graphene’s exceptional properties, such as its extremely high electron mobility and thermal conductivity, make it an excellent conductor, ideal for applications requiring efficient heat dissipation \cite{bolotin2008ultrahigh, balandin2008superior}. At small scales, graphene exhibits ballistic transport, allowing electrons to move with minimal resistance, which is crucial for developing ultra-fast electronic devices \cite{ando2005theory}. Graphene nanoribbons (GNRs) inherit many of graphene’s properties but offer additional tunability due to their reduced dimensions and edge effects. GNRs can have a tunable bandgap, making them promising candidates for semiconductor and transistor applications \cite{son2006energy}. Their electronic properties vary with edge shape: armchair-edged GNRs can be semiconducting with a width-dependent bandgap, while zigzag-edged GNRs often exhibit metallic behavior and magnetic edge states \cite{ezawa2006peculiar}. GNRs also maintain graphene's exceptional mechanical properties, including high tensile strength and flexibility, making them suitable for composite materials and nanoscale mechanical systems \cite{geim2007rise}. Additionally, GNRs can exhibit enhanced optical absorption, particularly in the infrared range, making them ideal for optoelectronic devices like photodetectors and sensors \cite{zhu2010graphene}. Their high surface-to-volume ratio and tunable electronic properties make GNRs highly sensitive to environmental changes, making them excellent candidates for nanoscale sensors in detecting gases, chemicals, or biomolecules \cite{huang2007substrate}.

\vspace{0.10cm}
\setlength{\parindent}{0pt}

Twisted bilayer graphene (TBG) refers to a structure composed of two layers of graphene stacked on top of each other with a slight rotational misalignment. One of the most remarkable discoveries in TBG is the emergence of superconductivity at the so-called "magic angle". At low temperatures, TBG can exhibit zero electrical resistance, a hallmark of superconductivity \cite{Cao2018}. This superconductivity is unconventional, likely arising from strong correlations between electrons rather than the traditional phonon-mediated mechanism found in conventional superconductors \cite{Yankowitz2019}. TBG can host topologically nontrivial phases, such as quantum anomalous Hall states, where the system exhibits quantized Hall conductivity without an external magnetic field \cite{Serlin2020}. These topological phases are robust against certain types of disorder and have potential applications in quantum computing and spintronics \cite{Bistritzer2011, Song2018}. Strain in TBG can further modify its electronic properties. By applying mechanical strain, researchers can alter the moiré pattern and band structure, offering another degree of control over the material's behavior \cite{Carr2018}. Twisted graphene structures can exhibit unique optical properties, including tunable absorption and emission spectra, which are influenced by the twist parameter and resulting moiré pattern \cite{Moon2013,Moon2014}. These properties make twisted graphene of interest for applications in photonics and optoelectronics \cite{Cao2018, Andrei2020}. From a theoretical perspective, the properties of TGNRs are often investigated through the massless one-body Dirac-like equations, even in the presence of electromagnetic fields \cite{kwan2020twisted,background}. The dynamics of Weyl pairs in magnetized TGNRs involves a complex interplay between twisting-induced band structure modifications and the effects of an applied magnetic field. The twist parameter, affects the degree of structural modification and the resulting electronic properties. As the twist parameter changes, the electronic bands can shift, leading to variations in energy levels and spatial distributions of such pairs. Additionally, the presence of a magnetic field can induce cyclotron orbits and alter the density of states, further impacting the behavior of Weyl pairs and charge carriers. However, these systems require investigation through fully covariant Dirac-Coulomb type two-body equations, especially when such driving forces are acting on the system \cite{guvendi-plb2}. Understanding these interactions requires a comprehensive theoretical approach. By embedding a curved surface into a flat Minkowski space-time \cite{wormhole} and deriving a fully-covariant many-body Dirac equation, we can model the system. Solving the resulting equation analytically allows for examining how different parameters, such as the twist parameter and magnetic field, affect the dynamics of quasi-particles. This exploration not only provides insights into the fundamental physics of Weyl fermions in twisted graphene but also has practical implications for designing advanced electronic and spintronic devices. The ability to control and manipulate Weyl pairs through structural and magnetic modifications opens up new possibilities for developing technologies that leverage the unique properties of these quasiparticles. However, significant challenges arise in many-body equations, particularly within the relativistic framework. A complete one-time many-body Dirac equation represents a fundamental advancement in the theoretical description of fermions, offering a more precise and rigorous framework compared to phenomenologically established many-body equations \cite{breit}. This equation provides a fundamental description of a system of $f\overline{f}$ pairs. Unlike phenomenological models, which often rely on empirical observations and approximations, this equation must be derived from first principles based on the two-body Dirac equation for bi-local fields \cite{bs}. These equations incorporate the intrinsic spin and relativistic effects of particles, which are essential for accurately describing systems where relativistic effects are significant \cite{barut}. Phenomenological many-body equations often simplify or approximate relativistic effects, focusing on non-relativistic or semi-relativistic regimes. A complete one-time many-body Dirac equation, however, fully accounts for relativistic corrections and spin interactions, providing a more accurate description of phenomena where relativistic effects are non-negligible, such as in high-energy physics or in materials with strong couplings. Such a fully covariant many-body Dirac equation was derived from quantum electrodynamics through the action principle \cite{barut}. This equation is indeed a complete equation, taking into account retardation effects appropriately and including spin algebra spanned by the Kronecker product of Dirac matrices \cite{nuri}. Additionally, this equation includes the most general electric and magnetic potentials, making it appropriate for studying the dynamics of many-body systems under the effects of external electromagnetic fields within flat \cite{g,g0} or curved spaces \cite{g1,g2,g3,g4,semra,proc,PLB2025}. This is crucial for a full understanding of many-body systems and provides an opportunity to predict phenomena that might not be captured by simpler or phenomenological models.

\vspace{0.10cm}
\setlength{\parindent}{0pt}

\begin{figure}
\centering
\includegraphics[scale=0.40]{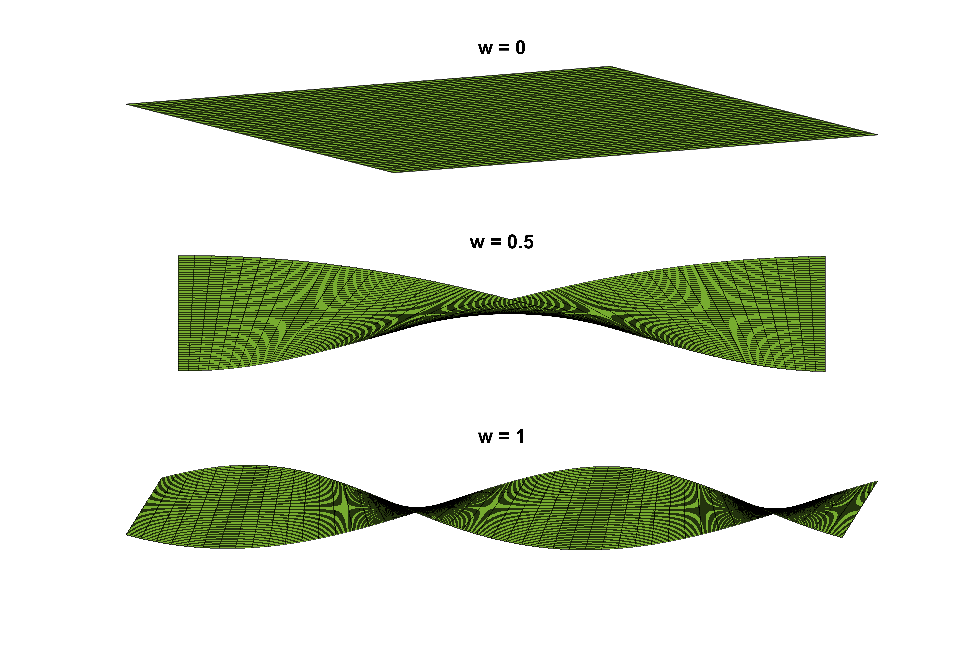}
\caption{\footnotesize Visualization of three graphene NRs with different twist parameters: \(w = 0\) (no twist), \(w = 0.5\), and \(w = 1\). The variations in the twist parameter \(w\) illustrate the effect of chirality on the ribbon geometry. The simulations are performed for a ribbon length \(L = 30\) and width \(D = 10\), emphasizing how the twist modulates the overall shape and spatial configuration of the NRs.}
\label{fig:2}
\end{figure}

It is well established that low-energy electronic excitations in flat graphene are accurately described by the massless two-dimensional Dirac equation. When graphene-based materials acquire curved geometries, non-trivial intrinsic curvature effects arise and can be systematically incorporated by extending this Dirac-like framework to its curved space formulation. While most theoretical investigations focus on solving the one-body Dirac equation \cite{background,HGNR,s5} due to the inherent complexity of many-body dynamics in curved space-times, the presence of an external magnetic field acting on charge carriers in helicoidal GNRs necessitates a simultaneous treatment of Weyl pairs. In this work, we consider a helicoidal GNR as a continuous and distortion-free structure, neglecting lattice discreteness and variations in hopping parameters (see also \cite{sugg-1}), and investigate the behavior of low-energy electronic excitations in such a magnetized helicoidal background. We focus on helicoidal GNRs, a distinctive class of low-dimensional materials in which a graphene strip undergoes continuous helical deformation, resulting in a screw-like geometry, and aim to obtain exact analytical solutions that reveal twist-induced effects on Weyl pairs by employing a fully covariant two-body Dirac equation. The manuscript is structured as follows: Section \ref{sec2} outlines the geometry of the magnetized helicoidal space-time under a uniform magnetic field aligned along the x-axis (see Figure \ref{fig:2}); Section \ref{sec3} derives a non-perturbative wave equation describing the relative motion of Weyl pairs; Section \ref{sec4} presents the exact solutions; Section \ref{sec5} provides the calculation of the polarization function; and Section \ref{sec6} concludes with a comprehensive summary and discussion of the results.

\section{\mdseries{Revisiting the magnetized helicoidal background}} \label{sec2}

The helicoidal background characterizing the TGNRs can be described using the parametrization \cite{background} (also see \cite{s5}):
\begin{eqnarray}
\vec{r}\left(u,\nu\right)=\nu\hat{i}+u \, \cos(\omega v)\hat{j}+u \, \sin(\omega v)\hat{k},
\end{eqnarray}
where \( v \in \left[-\frac{L}{2}, \frac{L}{2}\right] \) and \( u \in \left[-\frac{D}{2}, \frac{D}{2}\right] \). Here, \( D \) denotes the width of the NR, and \( L \) represents the total length of the ribbon, which is oriented along the \( x \)-axis. The twist parameter \( w = \frac{2\pi}{L}m \), where \( m \) is the number of \( 2\pi \) twists, is a real parameter with units of inverse length and governs the chirality of the surface. From this parametrization, the following expressions can be derived (for further details, refer to \cite{background}):
\begin{equation*}
\begin{split}
&dx = dv,\\
&dy = \frac{\partial y}{\partial u} du + \frac{\partial y}{\partial v} dv\Rightarrow \cos(w \nu) du - u w \sin(w v) dv,\\
&dz = \frac{\partial z}{\partial u} du + \frac{\partial z}{\partial v} dv\Rightarrow \sin(w v) du + u w \cos(w \nu) dv,\\
\end{split}
\end{equation*}
Accordingly, we have
\begin{equation*}
dx^2 + dy^2 + dz^2\Rightarrow du^2 + \left(1 +u^2 w^2\right)dv^2.
\end{equation*}
Also, by trivially projecting the temporal coordinate ($t$) from flat space-time, where the helicoidal ribbon lives, we can describe the TGNRs (i.e.,helicoidal surface) through the following ($2+1$)-dimensional curved space-time metric \cite{background}
\begin{eqnarray}
ds^2=c^2dt^2-du^2-\chi\left(u\right)dv^2, \label{eq1}
\end{eqnarray}
where $\chi\left(u\right)=1+w^2u^2$, and $c$ is the speed of light in vacuum. Referring to the line element Eq. (\ref{eq1}), one can derive the covariant metric tensor
\begin{equation*}
g_{\mu \nu}=\text{diag}\left(c^2, -1, -\chi\left(u\right)\right)
\end{equation*}
The space-independent (free) Dirac matrices are given in terms of the Pauli spin matrices ($\sigma^{x},\sigma^{y},\sigma^{z}$) as follows: $\gamma^{0}=\sigma^{z}, \gamma^{1}=i\sigma^{x}, \gamma^{2}=i\sigma^{y}$, in accordance with the signature ($+,-,-$) of the line element \cite{s2,s3,s4}. The spinorial affine connections for the Dirac field, denoted by \(\Gamma_{\lambda}\), are given by the relation: $\Gamma_{\lambda} = \frac{1}{4} g_{\mu \tau} \left[e^{k}_{\nu,\lambda} e^{\tau}_{k} - \Gamma_{\nu \lambda}^{\tau} \right]  \mathcal{S}^{\mu \nu}$, where \(_{,\lambda}\) indicates the derivative with respect to \(x^{\lambda}\). Here, \(\Gamma_{\nu \lambda}^{\tau}\) represents the Christoffel symbols, which are defined as: $\Gamma_{\nu \lambda}^{\tau} = \frac{1}{2} g^{\tau \epsilon} \left[\partial_{\nu} g_{\lambda \epsilon} + \partial_{\lambda} g_{\epsilon \nu} - \partial_{\epsilon} g_{\nu \lambda} \right]$,
  \(e^{\tau}_{k}\) denotes the inverse tetrad fields, and \(\mathcal{S}^{\mu \nu}\) symbolizes the spin operators, defined as:
 $\mathcal{S}^{\mu \nu} = \frac{1}{2} \left[\gamma^{\mu}, \gamma^{\nu} \right]$ \cite{s2,s3,s4}. Here, Greek indices refer the coordinates within the curved space-time and Latin indices stand for the coordinates of flat Minkowski space-time. The tetrad fields (and their inverses 
 $e^{\mu}_{k}$) can be obtained as follows
\begin{flalign*}
e^{k}_{\mu}=\left(\begin{array}{ccc}
c& 0 & 0\\
0& 1 & 0\\
0 & 0 & \sqrt{\chi(u)}
\end{array}\right), e^{\mu}_{k}=\left(\begin{array}{ccc}
\frac{1}{c}& 0 & 0\\
0& 1 & 0\\
0  & 0 & \frac{1}{\sqrt{\chi(u)}}
\end{array}\right)
\end{flalign*}
since $g_{\mu\nu}=e^{k}_{\mu}e^{l}_{\nu}\eta_{kl}$ and $e^{\mu}_{k}=g^{\mu\nu}e^{l}_{\nu}\eta_{kl}$ where $\eta_{kl}$ is the Minkowski tensor, $\eta_{kl}=\textrm{diag}\left(1,-1,-1\right)$ \cite{s2,s3,s4}.  Accordingly, we have $\gamma^{t}= \frac{1}{c}\sigma^{z}, \gamma^{u}= i \sigma^{x}, \gamma^{v}=\frac{1}{\sqrt{\chi(u)}}i \sigma^{y}$, where $i=\sqrt{-1}$. Also, non-zero components of the Christoffel symbols are found as follows: $\Gamma_{vv}^{u}=-u w^2, \Gamma_{v u}^{v}=\Gamma_{u v}^{v}=\frac{u w^2}{\chi(u)}$. Also, non-zero component of the spinorial affine connection is obtained as $\Gamma_{v}=\frac{u w^2}{2\sqrt{\chi(u)}} i \sigma^{z}$. Now let us derive the electromagnetic potential. Using the parametrization \cite{background}:
\begin{equation*}
\begin{pmatrix}
x \\
y \\
z
\end{pmatrix}
=
\begin{pmatrix}
v \\
u \cos (w v) \\
u \sin (w v)
\end{pmatrix},
\end{equation*}
we perform the standard procedure to write the new bases as:
\begin{equation*}
\hat{e}_i = \frac{\partial_i \vec{r}(u, v)}{|\partial_i \vec{r}(u, v)|},  (i = u, v).
\end{equation*}
Accordingly, we obtain:
\begin{equation*}
\begin{split}
&\hat{e}_u = \cos (w v) \hat{j} + \sin (w v) \hat{k},\\
&\hat{e}_v = \frac{\hat{i} - w u \sin (w v) \hat{j} + w u \cos (w v) \hat{k}}{\sqrt{1 + u^2 w^2}},
\end{split}
\end{equation*}
by using
\begin{equation*}
\begin{split}
&\frac{\partial \vec{r}(u, v)}{\partial u} = \cos (w v) \hat{j} + \sin (w v) \hat{k},\\
&\frac{\partial \vec{r}(u, v)}{\partial v} = \hat{i} - w u \sin (w v) \hat{j} + w u \cos (w v) \hat{k}.
\end{split}
\end{equation*}
Next, we perform the inverse procedure to express the Cartesian partial derivatives in terms of the partial derivatives \(\partial_i\) \cite{background}
\begin{equation*}
\begin{split}
&\partial x = \frac{\partial}{\partial v},\\
&\partial y = \cos (w v) \frac{\partial}{\partial u} - \frac{\sin (w v)}{w u} \frac{\partial}{\partial v},\\
&\partial z = \sin (w v) \frac{\partial}{\partial u} + \frac{\cos (w v)}{w u} \frac{\partial}{\partial v}.
\end{split}
\end{equation*}
Then, using the bases \(\hat{e}_u\) and \(\hat{e}_v\), we determine the spatial components of the electromagnetic vector potential \(\vec{\mathcal{A}}\):
\begin{equation*}
\vec{A} = \mathcal{A}_u \hat{e}_u + \mathcal{A}_v \hat{e}_v = \mathcal{A}_x \hat{i} + \mathcal{A}_y \hat{j} + \mathcal{A}_z \hat{k},
\end{equation*}
where
\begin{equation*}
\begin{split}
&\mathcal{A}_x = \frac{\mathcal{A}_v}{\sqrt{1 + u^2 w^2}},\\
&\mathcal{A}_y = \mathcal{A}_u \cos (w v) - \frac{\mathcal{A}_v}{w u} \sin (w v) \frac{1}{\sqrt{1 + u^2 w^2}},\\
&\mathcal{A}_z = \mathcal{A}_u \sin (w v) + \frac{\mathcal{A}_v}{w u} \cos (w v) \frac{1}{\sqrt{1 + u^2 w^2}}.\\
\end{split}
\end{equation*}
Now, let us consider a uniform magnetic field \(\mathcal{B}_0\) aligned along the axis \((\hat{i})\) of the helicoid. To obtain this, we need to calculate:
\begin{equation*}
\mathcal{B}_0 = \frac{\partial \mathcal{A}_z}{\partial y} - \frac{\partial \mathcal{A}_y}{\partial z}.
\end{equation*}
By choosing \(\mathcal{A}_u = 0\), we obtain the component of the vector potential:
\begin{equation}
\mathcal{A}_v = \frac{\mathcal{B}_0}{2 w} \sqrt{1 + u^2 w^2},\label{MF}
\end{equation}
which represents a uniform magnetic field aligned along the $x$-axis of the helicoid. This indicates that the momentum operator must be adjusted as follows:
\begin{equation*}
\partial_{v} \rightarrow \partial_{v} + i \frac{e \mathcal{A}_v}{2 \hbar c w} \sqrt{1 + u^2 w^2},
\end{equation*}
where \(e\) is the electric charge of the particle and \(\hbar\) is the reduced Planck constant, within each Dirac Hamiltonian (see \cite{guvendi-plb2}) to incorporate the effects of the uniform magnetic field. Consequently, the vector potential \(\mathcal{A}_v = \frac{\mathcal{B}_0}{2 w} \sqrt{1 + u^2 w^2}\) becomes singular at \(w = 0\) due to the \(1/w\) term. This singularity implies that the flat graphene results cannot be retrieved unless \(\mathcal{B}_0 = 0\), as the expression is otherwise undefined when \(w = 0\).

\section{\mdseries{Wave equation and exact results}}\label{sec3}

In this section, we will introduce the corresponding form of the fully-covariant two-body Dirac equation. We explicitly write the corresponding two-body Dirac equation for a Weyl pair ($m_1=m_2=0$) within the magnetized TGNR in the form of \(\hat{M}\Psi = 0\), where \(\hat{M}\) denotes (see also \cite{g3,g4,semra})
\begin{equation}
\begin{split}
&\gamma^{t^{f}}\otimes\gamma^{t^{\overline{f}}}\left[\partial_{t}^{f}+\partial_{t}^{\overline{f}} \right]+\gamma^{u^{f}}\partial_{u}^{f}\otimes \gamma^{t^{\overline{f}}}+ \gamma^{t^{f}}\otimes \gamma^{u^{\overline{f}}}\partial_{u}^{\overline{f}}\\
&+\gamma^{v^{f}} \otimes\gamma^{t^{\overline{f}}}\slashed{\partial}_{v}^{f}+\gamma^{t^{f}}\otimes \gamma^{v^{\overline{f}}}\slashed{\partial}_{v}^{\overline{f}}\\
&-\left[\gamma^{v^{f}}\Gamma_{v}^{f}\otimes \gamma^{t^{\overline{f}}}+\gamma^{t^{f}}\otimes \gamma^{v^{\overline{f}}}\Gamma_{v}^{\overline{f}} \right],\label{eq4}
\end{split}
\end{equation}
where $\gamma^{\mu}$ matrices indicate space-dependent Dirac matrices, the symbols $\otimes$ means Kronecker product, $\Psi(x_{\mu}^{f},x_{\mu}^{\overline{f}})$ is the bi-local spinor, $x_{\mu}^{f}$ and $x_{\mu}^{\overline{f}}$ are the space-time position vectors of the particles, and
\begin{equation*}
\begin{split}
&\slashed{\partial}_{v}^{f}= \partial_{v}^{f}+i\frac{e\mathcal{B}_{0}}{2\hbar  w}\sqrt{1+w^2u^{2}_{f}},\\
&\slashed{\partial}_{v}^{\overline{f}}= \partial_{v}^{\overline{f}}-i\frac{e\mathcal{B}_{0}}{2\hbar  w}\sqrt{1+w^2u^{2}_{\overline{f}}},
\end{split}
\end{equation*}
since \(e_{f}=-e_{\overline{f}}=e\) \cite{guvendi-plb2}. Following a comprehensive examination of the space-time interval in question, we express the space-time-dependent bi-spinor \(\Psi(t,r,R)\) in a factorized form, allowing us to decompose \(\Psi\) as \(e^{-i\frac{E}{\hbar} t} e^{i\vec{K} \cdot \vec{R}} \tilde{\Psi}(\vec{r})\), where
\begin{equation*}
\tilde{\Psi}(\vec{r}) = e^{i s v} (\psi_{1}\left(u\right), \psi_{2}\left(u\right), \psi_{3}\left(u\right), \psi_{4}\left(u\right))^T.
\end{equation*}
In this expression, \(E\) denotes the relativistic energy, while \(\vec{r}\) and \(\vec{R}\) represent the spatial position vectors for the relative motion and center of mass motion, respectively. The vector \(\vec{K}\) corresponds to the center of mass momentum, \(s\) indicates the total spin of the \(f\overline{f}\) pairs, and \(^T\) signifies the transpose operation on the \(u\)-dependent spinor. In accordance with the standard approach for analyzing two-body systems, we define the coordinates for relative motion and center of mass motion as follows \cite{guvendi-plb2}:
\begin{equation}
\begin{split}
&R_{x^{\mu}}=\frac{x^{\mu^{f}}}{2}+\frac{x^{\mu^{\overline{f}}}}{2},\quad r_{x^{\mu}}=x^{\mu^{f}}-x^{\mu^{\overline{f}}},\\
&x^{\mu^{f}}=\frac{1}{2}r_{x^{\mu}}+R_{x^{\mu}},\quad x^{\mu^{\overline{f}}}=-\frac{1}{2}r_{x^{\mu}}+R_{x^{\mu}},\\
&\partial_{x_{\mu}}^{f}=\partial_{r_{x^{\mu}}}+\frac{1}{2}\partial_{R_{x^{\mu}}},\quad \partial_{x_{\mu}}^{\overline{f}}=-\partial_{r_{x^{\mu}}}+\frac{1}{2}\partial_{R_{x^{\mu}}}, \label{eq4-}
\end{split}
\end{equation}
for a Weyl pair. It is important to note that \(\partial_{x^{\mu}}^{f} + \partial_{x^{\mu}}^{\overline{f}}\) can be simplified to \(\partial_{R_{x^\mu}}\). This simplification indicates that the system's evolution, governed by the relativistic energy \(E\), is related to the proper time, denoted by \(\partial_{R_{t}}\). By assuming that the center of mass remains fixed at the spatial origin, we can derive a set of equations describing the relative motion of the pair within the center of mass frame, where the total momentum (\(\hbar \vec{K}\)) is zero. For the relative motion of such spinless Weyl pairs, our model yields a system of equations
\begin{equation}
\begin{split}
&\tilde{ \mathcal{E}}\phi_{1}\left(u\right)-2\kappa\left(u\right)\phi_{3}\left(u\right)+2\hat{\lambda}\phi_{4}\left(u\right)=0,\\
&\tilde{ \mathcal{E}}\phi_{2}\left(u\right)=0,\\
&\tilde{ \mathcal{E}}\phi_{3}\left(u\right)-2\kappa\left(u\right)\phi_{1}\left(u\right)=0,\\
&\tilde{ \mathcal{E}}\phi_{4}\left(u\right)-2\hat{\lambda}\phi_{1}\left(u\right)=0,\label{eqset}
\end{split}
\end{equation}
where
\begin{equation*}
\begin{split}
&\tilde{ \mathcal{E}}=\frac{E}{\hbar c},\quad \kappa\left(u\right)=\tilde{\mathcal{B}}\sqrt{1+\frac{u^2w^2}{4}},\quad \tilde{\mathcal{B}}=\frac{e\mathcal{B}_{0}}{2\hbar w},\\
&\hat{\lambda}=\partial_{u}-\frac{uw^2}{\left(4+u^2w^2\right)}, 
\end{split}
\end{equation*}
and
\begin{equation*}
\begin{split}
&\phi_{1}\left(u\right)=\psi_{1}\left(u\right)+\psi_{4}\left(u\right),\quad \phi_{2}\left(u\right)=\psi_{1}\left(u\right)-\psi_{4}\left(u\right),\\
&\phi_{3}\left(u\right)=\psi_{2}\left(u\right)+\psi_{3}\left(u\right),\quad \phi_{4}\left(u\right)=\psi_{2}\left(u\right)-\psi_{3}\left(u\right).
\end{split}
\end{equation*} 

\section{\mdseries{Exact results}}\label{sec4}

The set of equations (\ref{eqset}) can be solved for \(\phi_{1}(u)\), yielding the following equation:
\begin{equation}
\hat{\lambda}\left[\hat{\lambda}\phi_{1}(u)\right]+\left[\tilde{ \mathcal{E}}^2/4-\kappa^{2}(u)\right]\phi_{1}(u)=0.\label{WE-1}
\end{equation}

\begin{figure}
\centering
\includegraphics[scale=0.55]{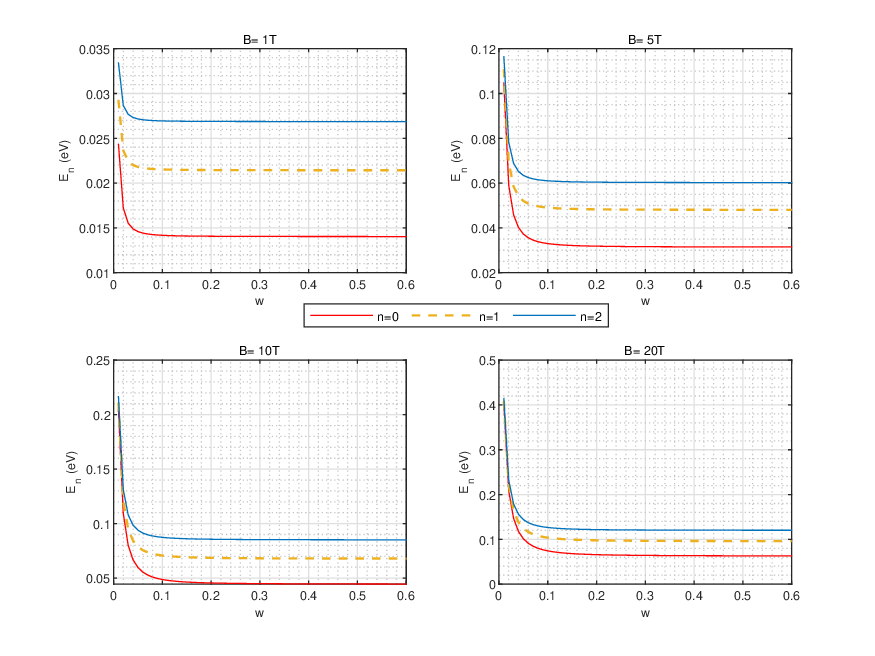}
\caption{\footnotesize Energy levels for different twist parameters and magnetic field strengths (Tesla).}
\label{fig:3}
\end{figure}

To find the exact solution of Eq. (\ref{WE-1}), we consider the ansatz \(\phi_{1}(u) = \frac{\phi(u)}{\sqrt{u}} \sqrt{4 + u^2 w^2}\). Substituting this into the wave equation, we get:
\begin{equation}
\ddot{\phi}(u) - \frac{1}{u} \dot{\phi}(u) + \left(\frac{\tilde{\mathcal{E}}^2 - 4\tilde{\mathcal{B}}^2}{4} + \frac{3}{4u^2} - \frac{u^2 w^2 \tilde{\mathcal{B}}^2}{4}\right)\phi(u) = 0,
\end{equation}
where the dot denotes differentiation with respect to \(u\). Introducing the new variable \(z = \frac{w \tilde{\mathcal{B}}}{2} u^2\) (with \(z \rightarrow 0\) as \(u \rightarrow 0\) and \(z \rightarrow \infty\) as \(u \rightarrow \infty\)), simplifies the equation to:
\begin{equation}
\ddot{\phi}(z) + \left[\frac{\tilde{\mu}}{z} - \frac{1}{4} + \frac{\frac{1}{4} - \tilde{\nu}^2}{z^2}\right]\phi(z) = 0,\label{Whit}
\end{equation}
where:
\begin{equation*}
\tilde{\mu} = \frac{\tilde{\mathcal{E}}^2 - 4\tilde{\mathcal{B}}^2}{8 \tilde{\mathcal{B}} w}, \quad \tilde{\nu} = \frac{1}{4}.
\end{equation*}
The solution to Eq. (\ref{Whit}) can be expressed using the Confluent Hypergeometric function:
\begin{equation}
\phi(z) = e^{-\frac{z}{2}} z^{\frac{1}{2} + \tilde{\nu}} \ _1F_1\left(\frac{1}{2} + \tilde{\nu} - \tilde{\mu}, 1 + 2\tilde{\nu}; z\right),\label{WF}
\end{equation}
around the regular singular point \(z = 0\) \cite{g3}. For \(\phi(z)\) to be finite and square-integrable, the Confluent Hypergeometric series must be truncated to a polynomial of order \(n \geq 0\). This requires the condition \(\frac{1}{2} + \tilde{\nu} - \tilde{\mu} = -n\) \cite{g3}, leading to the quantization condition (\(E \rightarrow E_{ns}\)) for the system. Consequently, by replacing \(c \rightarrow v_{F}\), the energy expression is:
\begin{equation}
E_{n} = 2 \frac{\hbar v_{F}}{\ell_{B}} \sqrt{n + \frac{3}{4} + \frac{1}{w^2 \ell_{B}^2}}, \label{spec-1}
\end{equation}
where \(v_{F} \approx \frac{c}{300}\) represents the Fermi velocity. This result applies to Weyl pairs within magnetized TGNRs. Here, \(\ell_{B} = \sqrt{\frac{\hbar}{e \mathcal{B}_{0}}}\) is the magnetic length, and \(w\) is in units of inverse length. The result indicates that the energy of such pairs approaches the Dirac point as \(\mathcal{B}_0 \rightarrow 0\). The effect of the twist parameter (\(w\)) on the energy levels is illustrated in Figure \ref{fig:3} for various magnetic field strengths (in Tesla). Notably, our exact results show that the energy spectra for Weyl pairs becomes (see also \cite{graphene}):
\begin{equation}
E_{\tilde{n}} = 2 \frac{\hbar v_{F}}{\ell_{B}} \sqrt{|\tilde{n}|}, \quad \tilde{n}=1,2,3...\label{spec-2}
\end{equation}
if \(\ell_{B} = 2/w\), and gives exact excited Landau levels of monolayer graphene. Moreover, there is no distinction between these energy levels \cite{graphene,L-weyl}. This implies that observations based on Landau levels for charge carriers in monolayer graphene may actually involve many-body effects, as demonstrated here. Figure \ref{fig:3} illustrates the influence of the magnetic field and twist parameter on energy values. As anticipated, the energy decreases with increasing twist parameter \( w \) up to a critical point determined by the magnetic field (or \( \ell_{B} \)). Beyond this point, the change in energy due to \( w \) becomes negligible for relatively large values of \( w \). This suggests that increasing \( w \) could lead to configurations where the system stabilizes by lowering its energy, which might explain the observed decrease in energy. Understanding how energy varies with these parameters can assist in designing systems or materials with desired properties, such as low-energy states for stability or specific magnetic behaviors.

\begin{figure}[ht]
    \centering
    \subfigure[]{\includegraphics[scale=0.15]{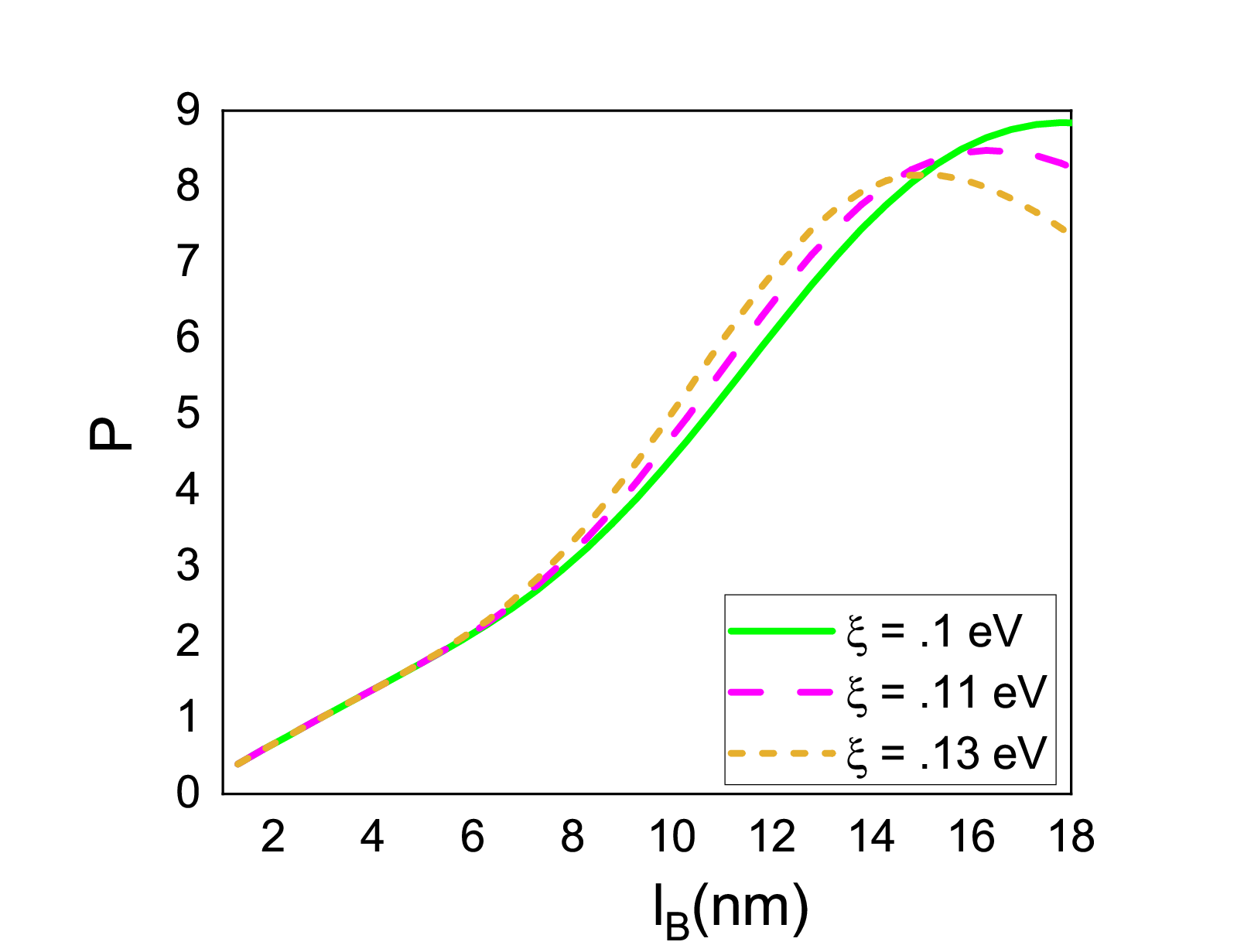}}\quad \subfigure[]{\includegraphics[scale=0.15]{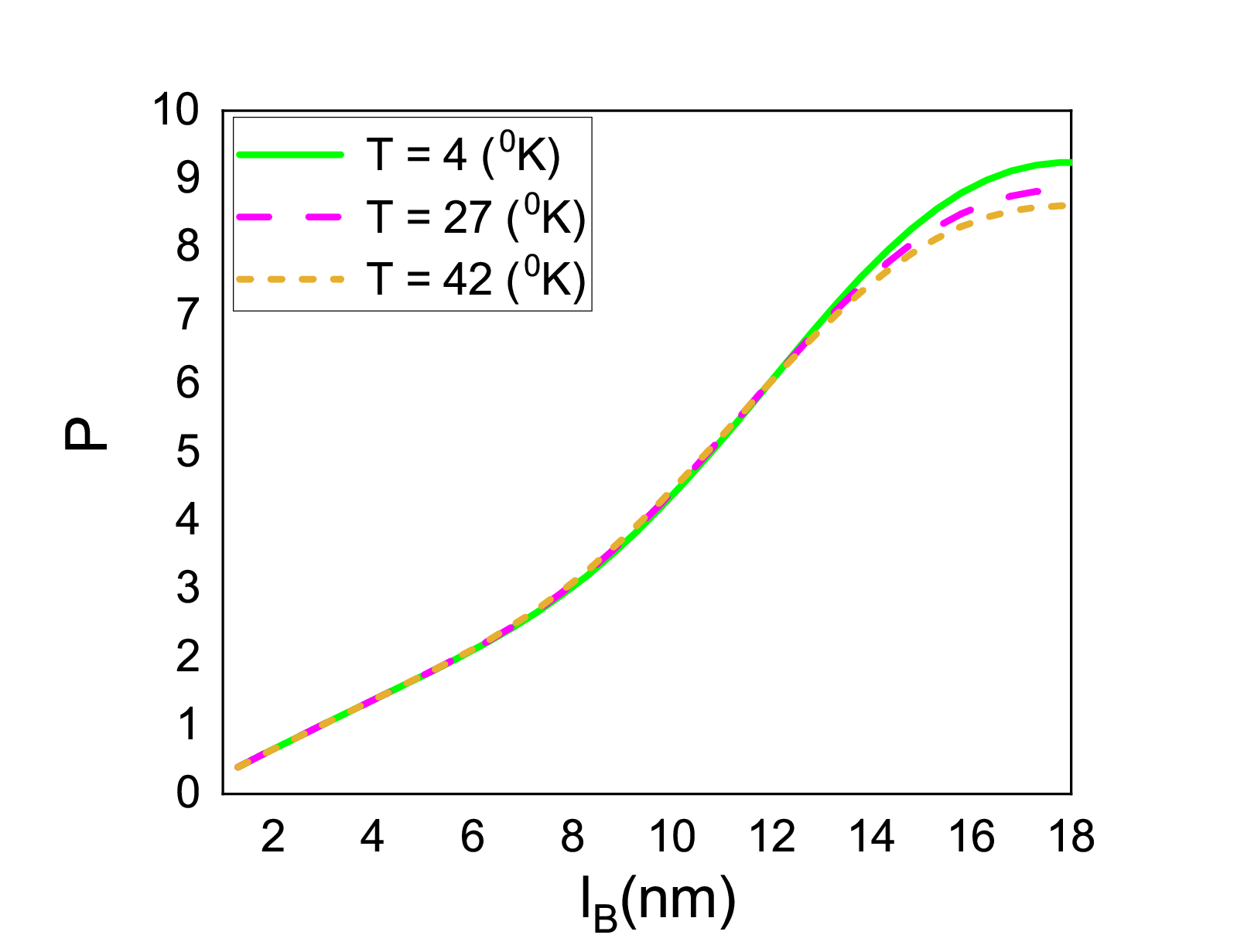}}\\
    \subfigure[]{\includegraphics[scale=0.15]{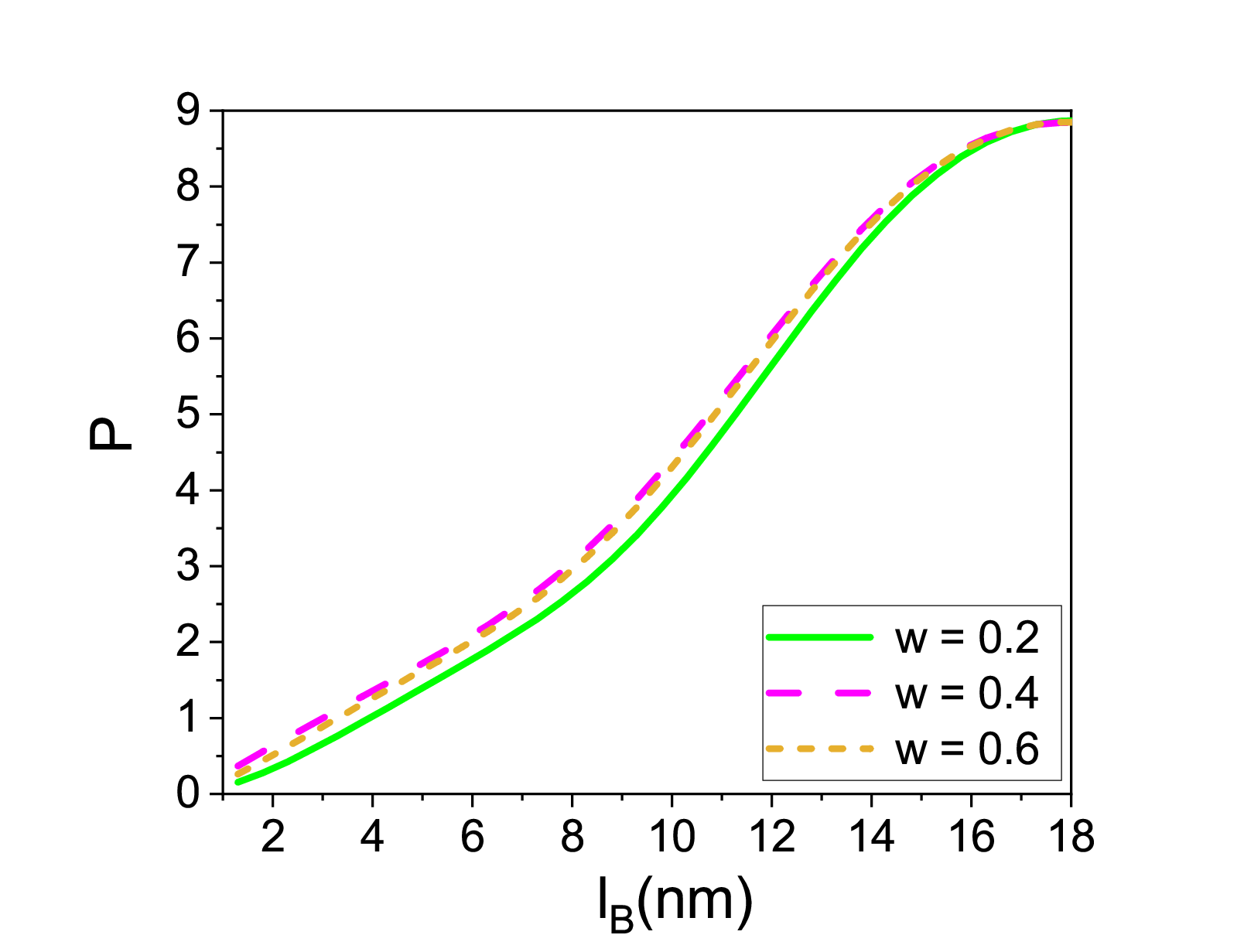}}\quad 
    \caption{\footnotesize  Polarization as a function of magnetic length.}
    \label{fig:5}
\end{figure}

\section{\mdseries{Polarization function}}\label{sec5}

\begin{figure}[ht]
    \centering
    \subfigure[]{\includegraphics[scale=0.15]{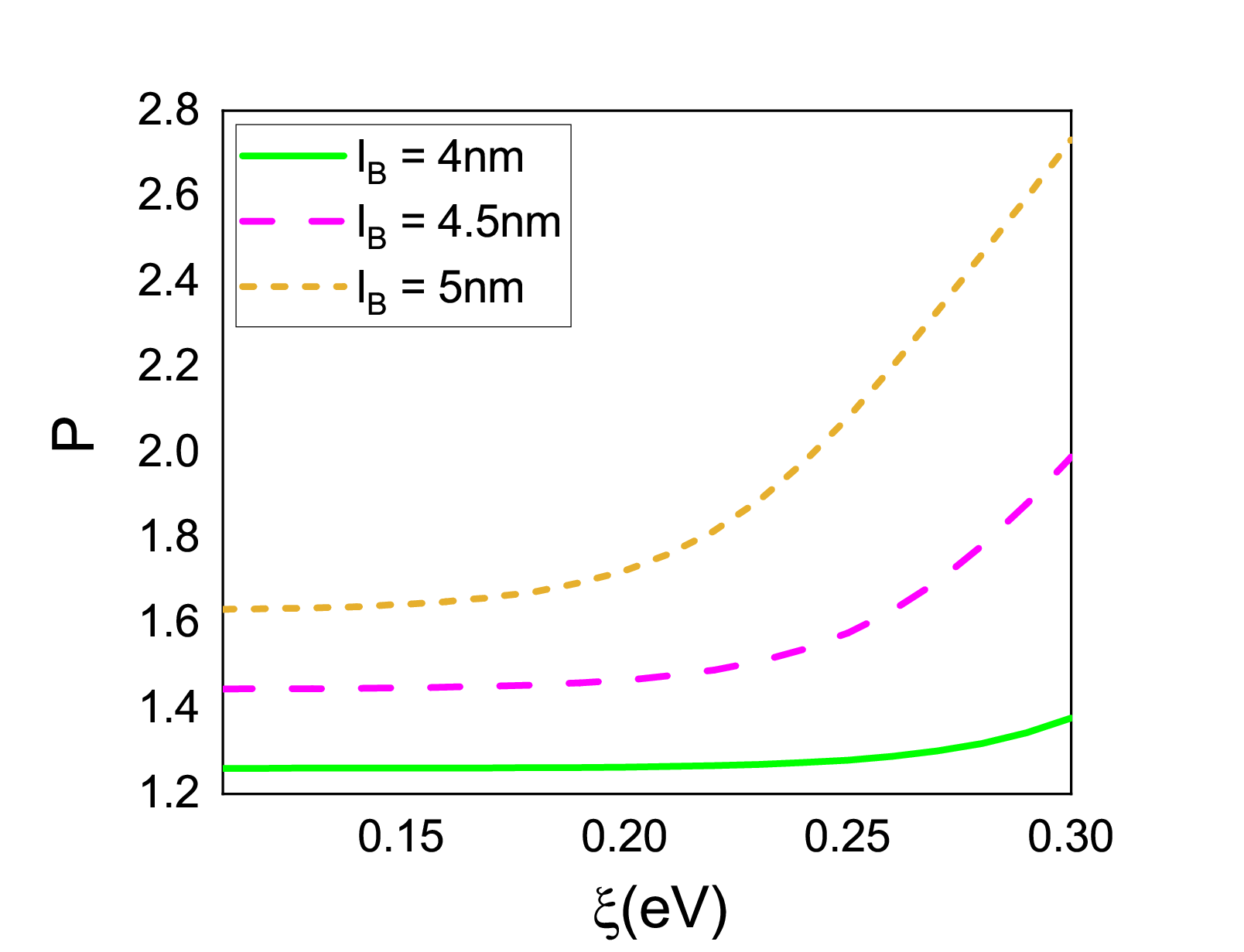}}
    \subfigure[]{\includegraphics[scale=0.15]{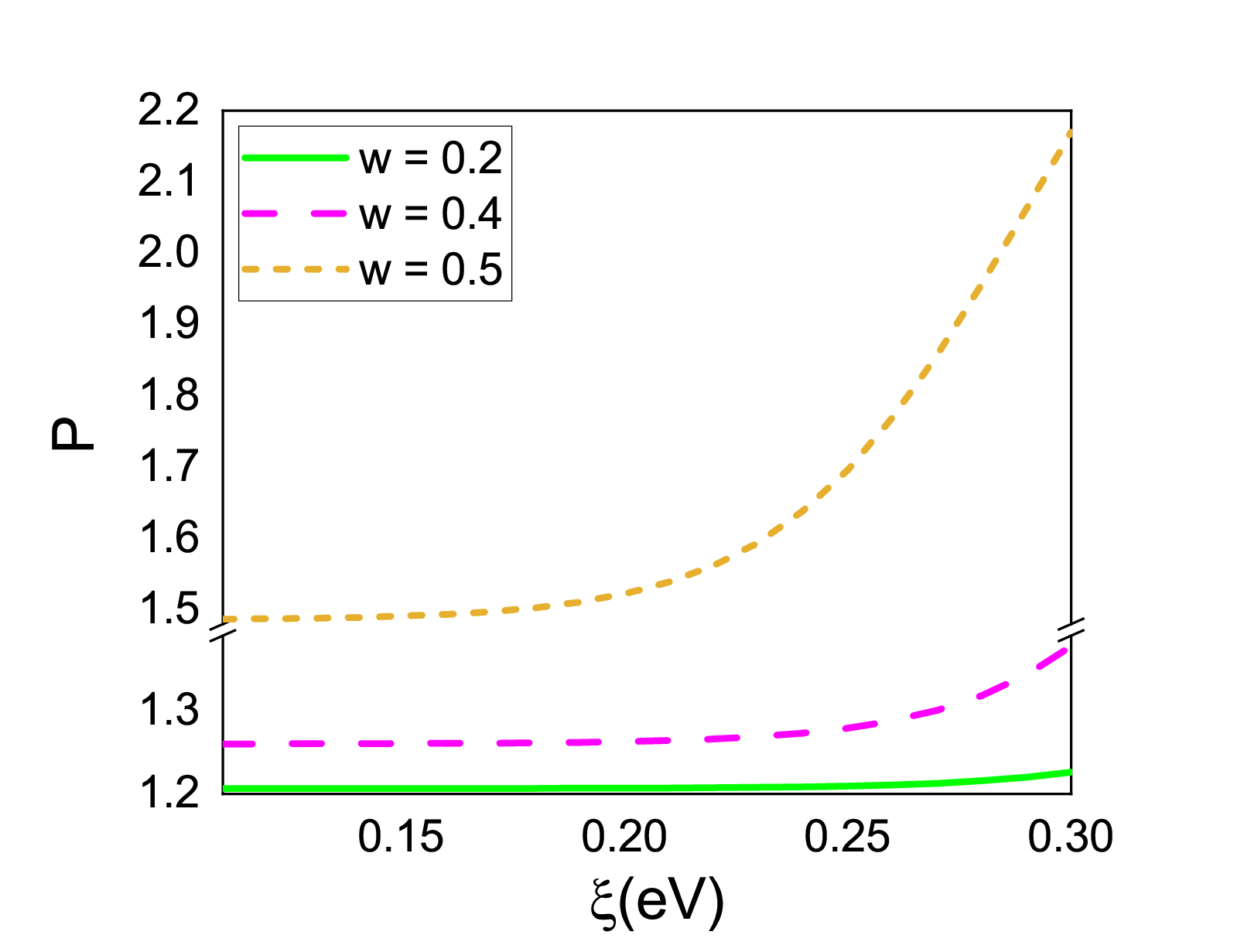}}\quad \subfigure[]{\includegraphics[scale=0.15]{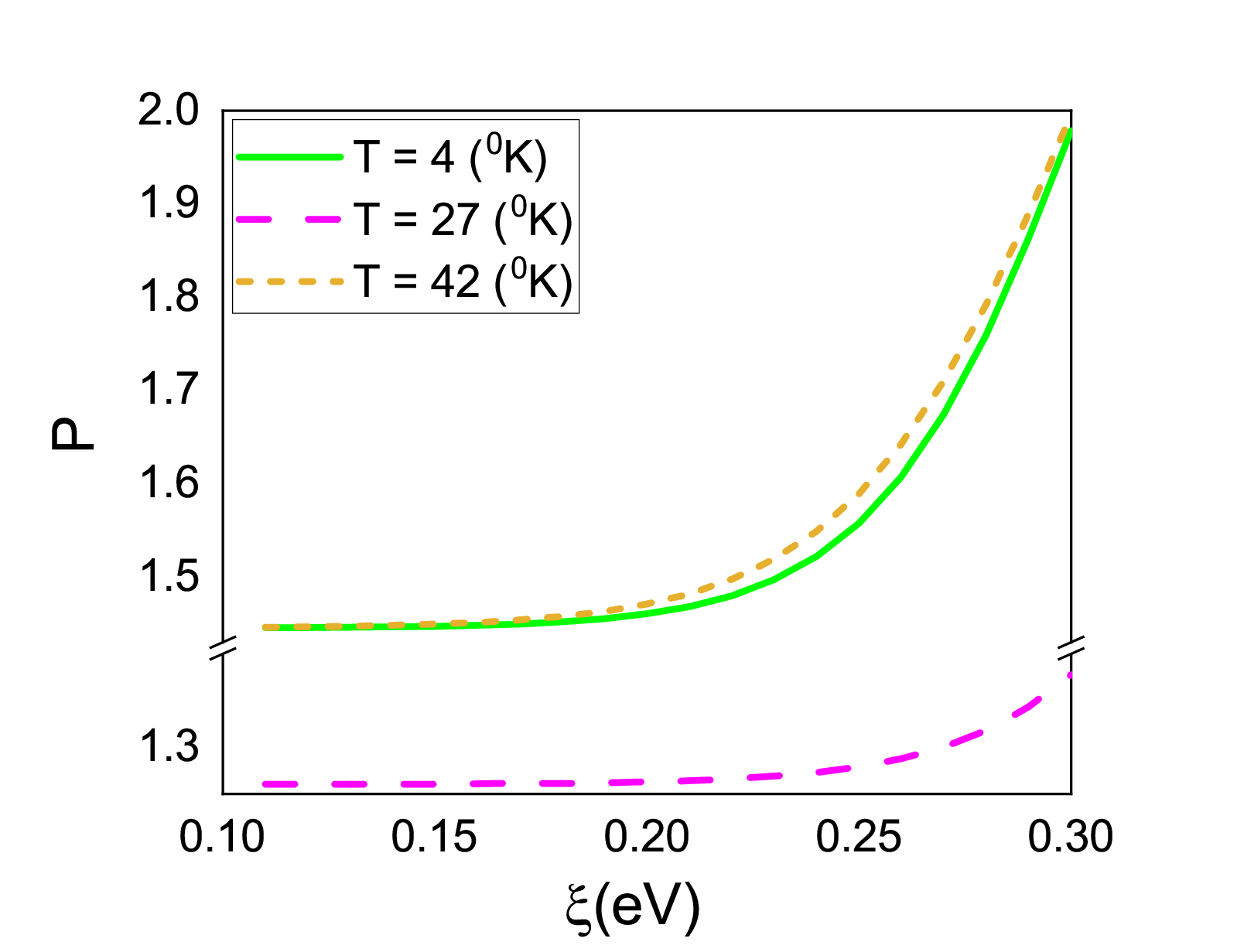}}
    \caption{\footnotesize  Polarization as a function of chemical potential.}
    \label{fig:6}
\end{figure}

The polarization function is a key concept in condensed matter physics and is related to the response of a system to external perturbations, such as electromagnetic fields. In the context of Weyl pairs in magnetized TGNRs, the polarization function describes how the electronic system within the NR responds to external fields and how this response is affected by the twisting and magnetic field.
The static polarization function can be defined as a coefficient of proportionality between the change of electron density and the screen potential and used for the calculation of the static dielectric function and conductivity limited by charged-impurity scattering. The polarization function \(P\) is given by \cite{P}
\begin{equation}
P = -g_v g_s \sum_{s, s', n, n'} (f_{s n} - f_{s' n'}) \frac{\left| \Phi_{s n}^\dagger \Phi_{s' n'} \right|^2}{E_{s n} - E_{s' n'}}.\label{Pol}
\end{equation}
Here, \(g_v\) and \(g_s\) are coupling constants or factors, namely \( g_v (= 2) \) is the valley degeneracy due to the \( K \) and \( K' \) Dirac points, and \( g_s (= 2) \) is the spin degeneracy. \(f_{s n}\) and \(f_{s' n'}\) represent occupation numbers or distribution functions, given by
\begin{equation}
f_{s n} = \frac{1}{1 + \exp \left( \frac{E_{s n} - \xi}{k_B T} \right)},\label{FD}
\end{equation}
\(\Phi_{s n}^\dagger\) and \(\Phi_{s' n'}\) are matrix elements or wavefunctions. \(E_{s n}\) and \(E_{s' n'}\) are energy levels associated with the states \((s, n)\) and \((s', n')\), respectively. In Eq. (\ref{FD}), \( \xi \) is the chemical potential, \( T \) is the temperature, and \( k_B \) is the Boltzmann constant. Here, \( s = +1 \) and \( s = -1 \) represent the conduction and valence bands, respectively. Our results allow us to observe the response of the \( P \) function to the parameters involved. Figure \ref{fig:5} presents the polarization as a function of \(\ell_{B}\) for three different values of chemical potential, temperature, and twist parameter. We observe that polarization decreases slightly with increasing temperature and chemical potential, whereas it increases with increasing magnetic length \(\ell_{B}\). Temperature, magnetic length, and chemical potential are crucial factors that can influence the polarization in a physical system. As the temperature increases, thermal energy agitates the particles in the system, which can affect their alignment or orientation. This can either enhance or reduce polarization, depending on the nature of the system. In systems influenced by magnetic fields, the energy levels (Landau levels) are quantized. The spacing and occupation of these levels depend on the magnetic length. Changes in these levels can directly affect the polarization, especially in low-dimensional systems. As \(\ell_{B}\) increases, the influence of the magnetic field on charge carriers diminishes. In many systems, a stronger magnetic field (smaller \(\ell_{B}\)) can align spins or charges more effectively, increasing polarization. The chemical potential determines the number of charge carriers in the system, which can directly influence polarization. The chemical potential essentially sets the Fermi level in a material. As it changes, different electronic states can become occupied or unoccupied, leading to changes in polarization. In certain systems, the chemical potential can influence ionic or electronic polarization by altering the distribution of ions or pairs within the material. This is particularly relevant in batteries, fuel cells, or electrolytic systems, where polarization effects can be directly controlled by adjusting the chemical potential. In designing materials and devices, understanding how these parameters control polarization can be crucial. Adjusting temperature, magnetic fields, or chemical potential can fine-tune sensitivity or selectivity. The same behavior can be observed in Figure \ref{fig:6}. These figures show that polarization increases for large chemical potential, especially for relatively high temperature and relatively large \(w\) values. Our results suggest that designing and controlling material properties for specific applications, such as in sensors, memory devices, or any technology, seems possible, especially by adjusting the magnetic field strength and the number of $2\pi$ twists on GNRs. 

\section{\mdseries{Summary and discussions}}\label{sec6}

In this study, we investigate Weyl pairs in magnetized helicoidal GNRs. By embedding a curved surface into flat Minkowski space-time, we derive a fully covariant two-body Dirac equation for the system, providing a non-perturbative wave equation for the relative motion of Weyl pairs. We solve this equation exactly, revealing how magnetic fields and structural twists influence the dynamics of spinless pairs in graphene NRs. The energy levels are given by:
\[
E_{n} = 2 \frac{\hbar v_{F}}{\ell_{B}} \sqrt{n + \frac{3}{4} + \frac{1}{w^2 \ell_{B}^2}},
\]
where \(\ell_{B} = \sqrt{\frac{\hbar}{e \mathcal{B}_0}}\) is the magnetic length and \(w\) is the twist parameter. As the magnetic field \(\mathcal{B}_0 \to 0\), the energy approaches the Dirac point. Figure \ref{fig:3} illustrates the influence of the twist parameter on energy levels at various magnetic field strengths. Our exact solutions show that the energy spectra of Weyl pairs are:
\[
E_{\tilde{n}} = \frac{\hbar v_F}{\ell_B / 2} \sqrt{|\tilde{n}|}, \quad \tilde{n}=1, 2, 3, \dots,
\]
for critical values of \(\ell_B = 2/w\), aligning with excited Landau levels of flat monolayer graphene, as reported in Refs. \cite{graphene, L-weyl}. This suggests no distinction between one-body and two-body states, implying that Landau level observations in graphene could reflect two-body effects. Figure \ref{fig:3} further demonstrates how the magnetic field and twist parameter affect energy. Energy decreases with increasing \(w\) up to a critical point, beyond which \(w\)’s effect on energy becomes negligible. This indicates that Weyl pairs in magnetized TGNRs exhibit distinctive properties due to curvature and twist-dependent magnetic fields, influencing their energy dispersion and electronic behavior. In contrast to Weyl pairs in topological insulators and Weyl semimetals, those in TGNRs can be tuned by adjusting the twist, with potential applications in quantum computing, and photonics.

\vspace{0.10cm}
\setlength{\parindent}{0pt}

The key advantages of Weyl pairs in TGNRs include tunable excitation properties and enhanced robustness in specific transport phenomena. However, challenges remain, such as precise twist angle control, structural defects, and modeling complexity. Our analytical framework improves the understanding of Weyl pairs in TGNRs, facilitating their potential applications. As the magnetic field approaches zero (\( \mathcal{B}_0 \to 0 \)), the energy approaches the Dirac point, suggesting a transition similar to that observed in Dirac semimetals. As \(w\) increases, the energy levels initially decrease, stabilizing the system at lower energies. This decrease becomes less pronounced as \(w\) increases further, marking the onset of a topologically saturated regime where energy levels become insensitive to further increases in \(w\). In this regime, the system reaches a stable configuration where the twist-induced effects and magnetic field balance, resulting in minimal energy variation. This stabilization is crucial for practical applications, such as the design of materials with controlled electronic behaviors, where stable energy configurations are essential. The topologically saturated regime may offer significant advantages in quantum computing, and photonics, where stable energy configurations are necessary for functionality.

\vspace{0.10cm}
\setlength{\parindent}{0pt}

Additionally, we analyze the polarization function as a function of chemical potential, temperature, twist parameter, and magnetic length. Figures \ref{fig:5} and \ref{fig:6} depict the polarization response to perturbations, examining how twisting and magnetic fields affect it. Figure \ref{fig:5} shows that polarization decreases slightly with higher temperature and chemical potential, while it increases with a larger magnetic length \(\ell_{B}\). Temperature, magnetic length, and chemical potential critically influence polarization. Increased temperature agitates particles, affecting their alignment and polarization. In magnetic systems, the quantization of Landau levels, influenced by the magnetic length, affects polarization, especially in low-dimensional systems. An increased \(\ell_{B}\) reduces the magnetic field's effect, while a stronger field (smaller \(\ell_{B}\)) can more effectively align spins or charges, enhancing polarization. The chemical potential, which determines the number of charge carriers, directly affects polarization by altering electronic state occupations. This is significant in systems like batteries or electrolytic systems, where polarization is controlled by adjusting the chemical potential. Figures \ref{fig:6} confirm that polarization increases with higher chemical potential, particularly at elevated temperatures and larger \(w\) values. Our results demonstrate that with precise tuning of both the magnetic field strength and the number of twist in GNRs, one can achieve exact control over material properties, which is essential for enhancing several applications.

\vspace{0.10cm}
\setlength{\parindent}{0pt}

To the best of our knowledge, there have been no direct experimental observations of Weyl pairs in TGNRs in the current literature. To validate our findings, experimental techniques such as transport measurements, Raman spectroscopy, scanning tunneling microscopy, and scanning tunneling spectroscopy could be employed. These methods would enable direct observations of the energy levels, local density of states (see also \cite{sugg-2,sugg-3}), and the modulation of electronic states in TGNRs under varying twist and magnetic field conditions. Our model offers a framework for guiding future experimental investigations, paving the way for exploring Weyl pairs in specifically engineered nanomaterials, with several potential applications.

%\section*{Acknowledgements}
%The authors thank ....

\section*{\small{CRediT authorship contribution statement}}

\textbf{Semra Gurtas Dogan}: Conceptualization, Methodology, Investigation, Writing – Review and Editing, Formal Analysis, Validation. \textbf{Kobra Hasanirokh}: Methodology, Data Curation, Investigation, Visualization, Writing – Review and Editing. \textbf{Omar Mustafa}: Conceptualization, Methodology, Investigation, Writing – Review and Editing. \textbf{Abdullah Guvendi}: Conceptualization, Methodology, Formal Analysis, Writing – Original Draft, Investigation, Visualization, Writing – Review and Editing, Project Administration.

\section*{\small{Data availability}}

The authors confirm that the data supporting the findings of this study are available within the article.

\section*{\small{Conflicts of interest statement}}

The authors have disclosed no conflicts of interest.

\section*{\small{Funding}}

This research has not received any funding.

\end{document}